# A Family of Estimators for Estimating the Population Mean in Simple random sampling under measurement errors

[1]Sachin Malik,   Jayant Singh and *[1]Rajesh Singh

[1]Department of Statistics, Banaras Hindu University

Varanasi-221005, India

[2]Department of Statistics, Rajasthan University, Jaipur

*Corresponding Author ( rsinghstat@gmail.com)

## Abstract

In this article we have suggested an improved estimator for estimating the population mean in simple random sampling using auxiliary information under the presence of measurement errors. The mean square error (MSE) of the proposed estimator has been derived under large sample approximation. Besides, considering the minimum case of the MSE equation, the efficient conditions between the proposed and existing estimators are obtained. These theoretical findings are supported by a numerical example.

## 1. Introduction

In survey sampling, the properties of the estimators based on data usually presuppose that the observations are the correct measurements on characteristics being studied. Unfortunately, this ideal is not met in practice for a variety of reasons, such as non response errors, reporting errors, and computing errors. When the measurement errors are negligible small, the statistical inferences based on observed data continue to remain valid. on the contrary when they are not appreciably small and negligible, the inferences may not be simply invalid and inaccurate but may often lead to unexpected, undesirable and unfortunate consequences (See Srivastava and Shalabh,2001).Some authors including Allen et al.(2003), Manisha and Singh (2001.2002), Shalabh (1997) and Singh and Karpe (2008,2009) have paid their attention towards the estimation of population mean $\mu_y$ of the study variable y using auxiliary information in the presence of measurement errors.

For a simple random sampling scheme, let $(x_i, y_i)$ be observed values instead of the true values $(X_i, Y_i)$ on two characteristics (x, y) respectively for the $i^{th}$ (i=1.2....n) unit in the sample of size n.

Let the measurement errors be

$$u_i = y_i - Y_i \tag{1.1}$$

$$v_i = x_i - X_i \tag{1.2}$$

Which are stochastic in nature with mean zero and variances $\sigma_u^2$ and $\sigma_v^2$ respectively, and are independent. Further, let the population means of (x, y) be $(\mu_x, \mu_y)$, population variances of (x, y) be $(\sigma_x^2, \sigma_y^2)$ and $\sigma_{xy}$ and $\rho$ be the population covariance and the population correlation coefficient between x and y respectively.(See Manisha and Singh (2002)).

In this chapter we have studied the behaviour of some estimators in presence of measurement error.

## 2. Some estimators in literature

The traditional ratio estimator is defined as

$$t_1 = \bar{y} \frac{\mu_x}{\bar{x}} \tag{2.1}$$

The bias and MSE of the estimator $t_1$ is given in Kumar et al. (2011), respectively, as

$$Bias(t_1) = \frac{1}{\mu_x}\left(R_m V_{xm} - V_{yxm}\right) \tag{2.2}$$

where $R_m = \frac{\mu_y}{\mu_x}$.

$$MSE(t_1) = \left[\frac{\sigma_y^2}{n} + \frac{R_m^2 \sigma_x^2}{n} - \frac{2R_m \rho_{yx}\sigma_y\sigma_x}{n}\right] + \frac{1}{n}\left[\frac{\mu_y^2}{\mu_x^2}\sigma_v^2 + \sigma_u^2\right] \tag{2.3}$$

MSE of the estimator $t_1$ can be re-written as

$$MSE(t_1) = M^*_{t_1} + M_{t_1} \tag{2.4}$$

where,

$M^*_{t_1} = \left[\dfrac{\sigma_y^2}{n} + \dfrac{R_m^2 \sigma_x^2}{n} - \dfrac{2R_m \rho_{yx} \sigma_y \sigma_x}{n}\right]$ is the mean squared error of the estimator $t_1$ without measurement error and

$M_{t_1} = \dfrac{1}{n}\left[\dfrac{\mu_y^2}{\mu_x^2}\sigma_v^2 + \sigma_u^2\right]$ is the contribution of measurement error in the estimator $t_1$.

Bahl and Tuteja (1991), suggested an exponential ratio type estimator for estimating $\overline{Y}$ as

$$t_2 = \bar{y}\exp\left(\dfrac{\mu_x - \bar{x}}{\mu_x + \bar{x}}\right) \tag{2.5}$$

The bias and MSE of the estimator $t_2$ is given by Kumar et al. (2011), respectively, as

$$Bias(t_2) = \dfrac{1}{\mu_x}\left(\dfrac{3}{8}R_m V_{xm} - \dfrac{1}{2}V_{yxm}\right) \tag{2.6}$$

$$MSE(t_2) = \dfrac{\sigma_y^2}{n}\left[1 - \dfrac{C_x}{C_y}\left(\rho - \dfrac{C_x}{4C_y}\right)\right] + \dfrac{1}{n}\left[\dfrac{\mu_y^2}{4\mu_x^2}\sigma_v^2 + \sigma_u^2\right] \tag{2.7}$$

MSE of the estimator $t_2$ can be expressed as

$$MSE(t_2) = M^*_{t_2} + M_{t_2} \tag{2.8}$$

where,

$M^*_{t_2} = \dfrac{\sigma_y^2}{n}\left[1 - \dfrac{C_x}{C_y}\left(\rho - \dfrac{C_x}{4C_y}\right)\right]$ is the MSE of the estimator $t_2$ without measurement error, and

$M_{t_2} = \dfrac{1}{n}\left[\dfrac{\mu_y^2}{4\mu_x^2}\sigma_v^2 + \sigma_u^2\right]$ is the contribution of measurement error in the estimator $t_2$.

Koyuncu and Kadilar (2010), suggested a regression type estimator $t_3$ as-

$$t_3 = \omega_1 \bar{y} + \omega_2 (\mu_x - \bar{x}) \tag{2.9}$$

where $\omega_1$ and $\omega_2$ are constants.

Kumar et al. (2011) have derived the bias and MSE of the estimator $t_3$ respectively, given by

$$Bias(t_3) = \mu_y (\omega_1 - 1) \tag{2.10}$$

$$MSE(t_3) = \mu_y^2 (\omega_1 - 1)^2 + \frac{1}{n}\omega_1^2 \sigma_y^2 + \frac{1}{n}\omega_2^2 \sigma_x^2 - \frac{2}{n}\omega_1 \omega_2 \rho \sigma_y \sigma_x + \frac{1}{n}\left(\omega_1^2 \sigma_u^2 + \omega_2^2 \sigma_v^2\right) \tag{2.11}$$

MSE of the estimator $t_3$ can be expressed as

$$MSE(t_3) = M_{t_3}^* + M_{t_3} \tag{2.12}$$

where,

$M_{t_3}^* = \mu_y^2 (\omega_1 - 1)^2 + \frac{1}{n}\omega_1^2 \sigma_y^2 + \frac{1}{n}\omega_2^2 \sigma_x^2 - \frac{2}{n}\omega_1 \omega_2 \rho \sigma_y \sigma_x$, is the MSE of the estimator $t_2$ without measurement error and

$M_{t_3} = \frac{1}{n}\left(\omega_1^2 \sigma_u^2 + \omega_2^2 \sigma_v^2\right)$, is the contribution of measurement error in the estimator $t_2$.

MSE of $t_3$ can be re-written as

$$MSE(t_3) = (\omega_1 - 1)^2 \mu_y^2 + \omega_1^2 (a_1) + \omega_2^2 a_2 + 2\omega_1 \omega_2 (-a_3) \tag{2.13}$$

where,

$a_1 = (V_{ym})$, $a_2 = (V_{xm})$, and $a_3 = (V_{yxm})$

Optimising MSE of the estimator $t_3$ with respect to $\omega_1$ and $\omega_2$, we get the optimum values of $\omega_1$ and $\omega_2$ as

$$\omega_1^* = \frac{b_3 b_4}{b_1 b_3 - b_2^2} \quad and \quad \omega_2^* = -\frac{b_2 b_4}{b_1 b_3 - b_2^2} \tag{2.14}$$

where,

$b_1 = \mu_y^2 + a_1, \quad b_2 = -a_3, \quad b_3 = a_2, \quad and \quad b_4 = \mu_y^2.$

Using these values of $\omega_1^*$ and $\omega_2^*$ from equation (2.14) into equation (2.13), we get the minimum MSE of the estimator $t_3$ as

$$MSE(t_3)_{min} = \left[ \mu_y^2 - \frac{b_3 b_4^2}{b_1 b_3 - b_2^2} \right] \qquad (2.15)$$

In the same way Bahl and Tuteja (1991), suggested an exponential product type estimator for estimating $\overline{Y}$ as

$$t_4 = \bar{y} \exp\left( \frac{\bar{x} - \mu_x}{\bar{x} + \mu_x} \right) \qquad (2.16)$$

$$Bias(t_4) = \frac{1}{\mu_x} \left( \frac{1}{8} R_m V_{xm} + \frac{1}{2} V_{yxm} \right) \qquad (2.17)$$

$$MSE(t_4) = \frac{\sigma_y^2}{n} \left[ 1 + \frac{C_x}{C_y} \left( \rho - \frac{C_x}{4C_y} \right) \right] + \frac{1}{n} \left[ \frac{\mu_y^2}{4\mu_x^2} \sigma_v^2 - \sigma_u^2 \right] \qquad (2.18)$$

$$MSE(t_4) = M_{t_4}^* + M_{t_4} \qquad (2.19)$$

where,

$M_{t_4}^* = \frac{\sigma_y^2}{n} \left[ 1 + \frac{C_x}{C_y} \left( \rho - \frac{C_x}{4C_y} \right) \right]$ is the mean squared error of $t_4$ without measurement error and

$M_{t_4} = \frac{1}{n} \left[ \frac{\mu_y^2}{4\mu_x^2} \sigma_v^2 + \sigma_u^2 \right]$ Is the contribution of measurement errors in $t_4$

Using the estimator given in (2.1), we suggest Singh and Vishvakarma (2005) estimator under measurement error as

$$t_5 = \bar{y} \left[ \alpha \frac{\mu_x}{\bar{x}} + (1-\alpha) \frac{\bar{x}}{\mu_x} \right] \qquad (2.15)$$

The bias and MSE of the estimator is given by

$$Bias(t_5) = \frac{1}{\mu_x}\left(\alpha R_m V_{xm} + V_{yxm} - 2\alpha V_{yxm}\right) \qquad (2.16)$$

$$MSE(t_5) = V_{ym} + V_{xm} R_m^2 \left[1 + 4\alpha^2 - 4\alpha\right] + 2V_{yxm} R_m \left[1 - 2\alpha\right] \qquad (2.17)$$

Where, $\alpha = \dfrac{R_m V_{xm} + V_{yxm}}{2 R_m V_{xm}}$ \hfill (2.18)

For the given value of $\alpha$ in equation (2.18) MSE of the estimator $t_5$ will be minimum.

## 3. Proposed estimator

Following Singh and Vishvakarma (2005), we propose a new family of estimators

$$t_p = q\left[\bar{y}\left(\frac{\mu_x}{\bar{x}}\right)^{m1} \exp\left(\frac{\mu_x - \bar{x}}{\mu_x + \bar{x}}\right)^{m2}\right] + (1-q)\left[\bar{y}\left(\frac{\bar{x}}{\mu_x}\right)^{m1} \exp\left(\frac{\bar{x} - \mu_x}{\bar{x} + \mu_x}\right)^{m2}\right] \qquad (3.1)$$

Where, q is a real constant to be determined such that the MSE of $t_p$ is a minimum and m1 and m2 are real constants such that m1+m2=1

Expanding equation (3.1) and subtracting $\mu_y$ from both sides then taking expectation of both the sides we get the bias of the estimator $t_p$ to the first degree of approximation

$$Bias(t_p) = V_{xm}\left[\frac{m_1(m_1+1)R_m}{2\mu_x} + \frac{m_1 R_m}{2\mu_x} + \frac{R_m}{8\mu_x}(1-m_1)\frac{R_m}{4\mu_x}\right]$$

$$+ V_{yxm}\left[\frac{m_1 R_m}{2\mu_x} + \frac{1}{2\mu_x} - \frac{2m_1 q}{\mu_x} - \frac{q(1-m_1)R_m}{4\mu_x}\right]$$

From (3.1), we have

$$(t_p - \mu_y)^2 = \left[\mu_y + k_1 - 2m_1 q k_2 R_m + k_2 R_m\left(m_1 - \frac{q}{2}\right)\right]^2 \qquad (3.2)$$

Taking expectation of both the sides of (3.2) the MSE to the first order of approximation will be

$$MSE(t_p) = \mu_y^2 + V_{ym} + R_m^2 V_{xm}\left[m_1^2 + q^2/4 + 4q^2 m_1^2 + 2q^2 m_1 - 4q m_1^2\right]$$

$$- R_m q V_{xm} m_1 - V_{yxm} R_m [4qm_1 + q] \quad (3.3)$$

Minimisation of (3.3) with respect to $q$ yields its optimum values as

$$q = \frac{4R_m m_1^2 V_{xm} + 2V_{xm} m_1 + V_{yxm}[4m_1 + 1]}{V_{xm} R_m [8m_1^2 + 1/2]}$$

**Remark 3.1:** For the different values of q and $m_1$, we can find out the estimators given in literature respectively.

## 4. Empirical Studies:

**Data statistics:** The data used for empirical study has been taken from Gujrati and Sangeetha (2007) -pg, 539.

Where, $Y_i$= True consumption expenditure,

$X_i$= True income,

$y_i$= Measured consumption expenditure,

$x_i$ = Mesured income.

From the data given we get the following parameter values

**Table 4.1:**

| N | $\mu_y$ | $\mu_x$ | $\sigma_y^2$ | $\sigma_x^2$ | $\rho$ | $\sigma_u^2$ | $\sigma_v^2$ |
|---|---|---|---|---|---|---|---|
| 10 | 127 | 170 | 1278 | 3300 | 0.964 | 36.00 | 36.00 |

**Table 4.2: Showing the MSE and PRE'S of the estimators with and without measurement errors**

| Estimators | MSE without meas. Error | Contribution of meas. Error in MSE | MSE with meas. Error | PRE'S |
|---|---|---|---|---|
| $\bar{y}$ | 127.800 | 3.600 | 131.400 | 100 |
| $t_1$ | 16.181 | 5.609 | 21.790 | 603.011 |
| $t_2$ | 142.357 | 5.132 | 147.489 | 86.650 |
| $t_3$ | 25.925 | 4.102 | 30.027 | 437.596 |
| $t_4$ | 178.524 | 6.025 | 184.549 | 69.249 |
| $t_5$ opt. | 8.972 | 4.932 | 13.904 | 944.943 |
| $t_P$ opt. | 8.332 | 4.572 | 13.904 | 944.943 |

**Conclusion:**

From the theoretical discussion and empirical study we conclude that the proposed estimator under optimum conditions performs better than other estimators considered in the article. The relative efficiency of various estimators are listed in Table 4.2.

**References**

Allen, J., Singh, H. P. and Smarandache, F. (2003): A family of estimators of population mean using multiauxiliary information in presence of measurement errors. International Journal of Social Economics 30(7), 837–849.

A.K. Srivastava and Shalabh (2001). Effect of measurement errors on the regression method of estimation in survey sampling. Journal of Statistical Research, Vol. **35**, No. 2, pp. 35-44.

Bahl, S. and Tuteja, R. K. (1991): Ratio and product type exponential estimator. Information and optimization sciences12 (1), 159-163.

Gujarati, D. N. and Sangeetha (2007): Basic econometrics. Tata McGraw – Hill.

Koyuncu, N. and Kadilar, C. (2010): On the family of estimators of population mean in stratified sampling. Pakistan Journal of Statistics. Pak. J. Stat. 2010 vol 26 (2), 427-443.

Kumar M., Singh R., Singh A.K. and Smarandache F. (2011): some ratio- type estimators under measurement errors. World applied science journal 14(2): 272-276.

Manisha and Singh, R. K. (2001): An estimation of population mean in the presence of measurement errors. Journal of Indian Society of Agricultural Statistics 54(1), 13–18.

Manisha and Singh, R. K. (2002): Role of regression estimator involving measurement errors. Brazilian journal of probability Statistics 16, 39- 46.

Shalabh (1997): Ratio method of estimation in the presence of measurement errors. Journal of Indian Society of Agricultural Statistics 50(2):150– 155.

Singh, H. P. and Karpe, N. (2008): Ratio-product estimator for population mean in presence of measurement errors. Journal of Applied Statistical Sciences 16, 49–64.

Singh, H. P. and Karpe, N. (2009): On the estimation of ratio and product of two populations means using supplementary information in presence of measurement errors. Department of Statistics, University of Bologna, 69(1), 27-47.

Singh, H. P. and Vishwakarma, G. K. (2005): Combined Ratio-Product Estimator of Finite Population Mean in Stratified Sampling. *Metodologia de Encuestas* 8: 35- 44.